\documentclass[
     ,draft            
   ]
   {aipproc}

\layoutstyle{8x11double}

\newcommand{\Mdot} {\dot{M}}
\newcommand{\Pdot} {\dot{P}}
\newcommand{\Mwd}  {M_{\rm wd}}
\newcommand{\Rwd}  {R_{\rm wd}}
\newcommand{\Msun} {{\rm M_{\odot}}}
\newcommand{\lax}{{\lower0.75ex\hbox{ $<$ }\atop\raise0.5ex\hbox{ $\sim$ }}}
\newcommand{\gax}{{\lower0.75ex\hbox{ $>$ }\atop\raise0.5ex\hbox{ $\sim$ }}}
\newcommand{\ion}[2]{#1~{\sc\uppercase\expandafter{\romannumeral #2}}}

\begin{document}


\title{
Optical, UV, and EUV Oscillations of SS~Cygni in Outburst\footnotemark{}
\textsuperscript{\footnotesize,}\footnotemark{}}

\author{Christopher W.~Mauche}{
   address={Lawrence Livermore National Laboratory,
   L-473, 7000 East Avenue, Livermore, CA 94550},
}

\begin{abstract}
I provide a review of observations in the optical, UV ({\it HST\/}),
and EUV ({\it EUVE\/} and {\it Chandra\/} LETG) of the rapid periodic
oscillations of nonmagnetic, disk-accreting, high mass-accretion rate
cataclysmic variables (CVs), with particular emphasis on the dwarf nova
SS~Cyg in outburst. In addition, I drawn attention to a correlation,
valid over nearly six orders of magnitude in frequency, between the
frequencies of the quasi-periodic oscillations (QPOs) of white dwarf,
neutron star, and black hole binaries. This correlation identifies the
high frequency quasi-coherent oscillations (so-called ``dwarf nova
oscillations'') of CVs with the kilohertz QPOs of low mass X-ray
binaries (LMXBs), and the low frequency and low coherence QPOs of CVs
with the horizontal branch oscillations (or the broad noise component
identified as such) of LMXBs. Assuming that the same mechanisms produce
the QPOs of white dwarf, neutron star, and black hole binaries, this
correlation has important implications for QPO models.

\end{abstract}

\maketitle


\addtocounter{footnote}{-1}
\footnotetext{Dedicated to my colleague and friend Janet Akyuz Mattei,
who this year celebrates her 30th year as director of the American
Association of Variable Star Observers.}
\addtocounter{footnote}{1}
\footnotetext{Based in part on observations with the NASA/ESA {\it
Hubble Space Telescope\/} obtained at the Space Telescope Science
Institute, which is operated by the Association of Universities for
Research in Astronomy, Incorporated, under NASA contract NA5-26555.}


\title{Optical, UV, and EUV Oscillations of SS~Cygni in Outburst}


\section{Introduction}

Rapid periodic oscillations have been studied in cataclysmic variables
(CVs; \cite{war95}) since the early 1970s in the optical and the early
1980s in the extreme ultraviolet  (EUV) and soft X-rays \cite{rob76,
pat81, war04}. (Nominally) nonmagnetic, disk-accreting, high
mass-accretion rate (``high-$\Mdot $;'' novalike variables and dwarf
novae in outburst) CVs manifest ``dwarf nova oscillations'' (DNOs) with
periods $P\approx 3$--30~s and high coherence ($Q\approx 10^4$--$10^6$),
as well as ``quasi-periodic oscillations'' (QPOs) with longer periods
(by a factor of $\approx 13$, see \S 3) and far lower coherence
($Q\approx 1$--$10$). In contrast, the oscillations of intermediate
polars/DQ~Her stars \cite{pat94}, the white dwarf analogues of
accretion-powered X-ray pulsars, have longer periods ($P= 33$--7190 s)
and far higher coherence ($Q\approx 10^{10}$--$10^{12}$). While the
pulsation periods of intermediate polars vary gradually ($|\Pdot |
<10^{-10}~\rm s~s^{-1}$), those of dwarf novae vary significantly during
dwarf nova outbursts, decreasing on the rising branch and increasing on
the declining branch of the outburst.

Although QPO research in compact binaries largely shifted to LMXBs in
the mid-1980s, the study of rapid oscillations in CVs continues to be of
value because of the many similarities between CVs and LMXBs, because
CVs are in many ways better laboratories in which to study accretion
processes (the mass-accretion rate varies systematically by three orders
of magnitude during dwarf nova outbursts, the peak luminosities are
well below the Eddington value, radiation pressure is unimportant, and
relativistic effects are insignificant), because CV oscillations can be
studied from the ground in the optical with modest-size telescopes, and
because CVs offer unique diagnostics of the oscillations in the optical,
UV, and EUV. Since the date of this meeting, Warner \cite{war04} has
supplied an excellent review of the oscillations of CVs, particularly
those in the optical. This contribution concentrates on observations
of oscillations in the optical, UV, and EUV flux of the best-studied CV,
the dwarf nova SS~Cyg. During outburst, oscillations have been detected
in its optical \cite{pat78, hor80, hil81, pat81}, UV ({\it HST\/}) [\S
4], and EUV/soft X-ray ({\it HEAO 1\/}, {\it EXOSAT\/}, {\it ROSAT\/},
and {\it EUVE\/}) \cite{cor80, cor84, jon92, mau96, tes97, mau01, mau02}
flux with periods ranging from 3~s to 11~s.

\begin{figure} 
\label{fig1}
\resizebox{0.95000\columnwidth}{!}{\includegraphics{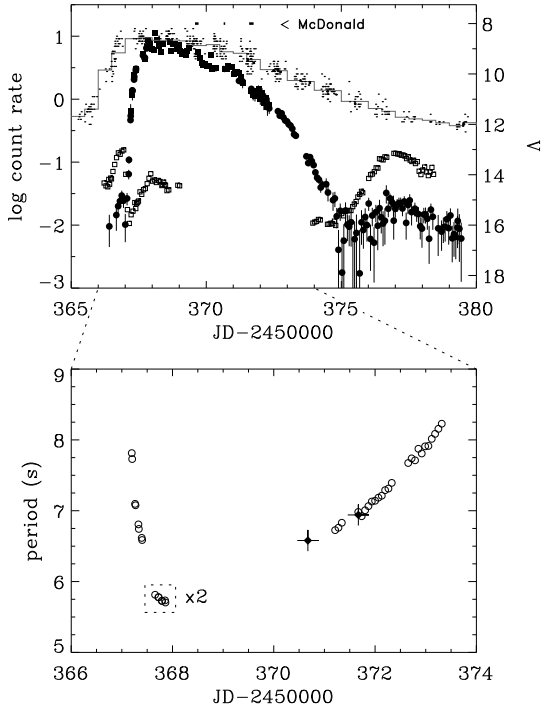}}
\caption{{\it Upper panel\/}: AAVSO optical, {\it EUVE\/}, and {\it
RXTE\/} light curves of the 1996 October outburst of SS~Cyg. {\it EUVE\/}
DS and SW measurements are shown respectively by the filled circles and
squares with error bars, {\it RXTE\/} PCA measurements (kindly supplied
by P.\ Wheatley) are shown by the open squares, individual AAVSO
measurements are shown by the small dots, and the half-day mean optical
light curve is shown by the histogram. Intervals of observations at
McDonald Observatory are indicated by the thick bars. {\it Lower panel\/}:
Oscillation period versus time. {\it EUVE\/} DS and McDonald Observatory
optical measurements are shown by the open circles and filled starred
diamonds, respectively. Points enclosed by the dotted box are plotted at
twice the observed periods.}
\end{figure}

\section{Simultaneous Optical, UV, and EUV Observations}

Mauche \& Robinson \cite{mau01} and Wheatley, Mauche, \& Mattei
\cite{whe03} have described simultaneous optical (AAVSO visual
magnitudes and McDonald Observatory 2.7 m telescope high-speed $UBVR$
photometry), EUV ({\it EUVE\/} DS and SW: $\lambda\approx 70$--120~\AA
), and X-ray ({\it RXTE\/} PCA: $E\approx 2$--15 keV) observations of a
narrow asymmetric outburst of SS~Cyg in 1996 October. The resulting
optical, EUV, and X-ray light curves are shown in the upper panel of
Figure~1. They can be understood in the context of the thermal-viscous
instability model of dwarf nova outbursts \cite{las01, sch03}. When the
disk surface density reaches some critical value first in the outer
disk, the disk plasma is heated locally and a heating wave is launched
though the disk, heating the disk plasma and increasing its viscosity.
As this material sinks toward the white dwarf, it converts its
gravitational potential energy into rotational kinetic energy and
radiation. This radiation comes out first in the optical and then in
the UV as hotter parts of the disk are activated (in steady state,
$T_{\rm disk}\propto r^{-3/4}$). When this material reaches the boundary
layer between the disk and the surface of the white dwarf, it converts
its prodigious rotational kinetic energy into radiation. This radiation
comes out first in X-rays because the boundary layer is initially
optically thin and hence quite hot (of order the virial temperature
$kT_{\rm vir}=G\Mwd m_{\rm p}/3\Rwd\sim 10$ keV), and then in the EUV
when the boundary layer becomes optically thick to its own radiation and
becomes relatively cool (of order the blackbody temperature $kT_{\rm
bb}=k[G\Mwd\Mdot/8\pi\sigma\Rwd ^3]^{1/4}\sim 10$ eV). The delay between
the rise of the optical and EUV/X-ray emission is a direct measure of
the time it takes for the heating wave to sweep through the disk. The
delays of approximately $1\frac{1}{2}$, $1\frac{1}{4}$, and $\frac{3}{4}$
days between the rise of the optical and the EUV flux of outbursts of
SS~Cyg, U~Gem, and VW~Hyi, respectively, are consistent with a heating
wave velocity $r_{\rm disk}/t_{\rm delay}\approx 3~\rm km~s^{-1}$
\cite{mau01c}.

\begin{figure} 
\label{fig2}
\resizebox{0.83125\columnwidth}{!}{\includegraphics{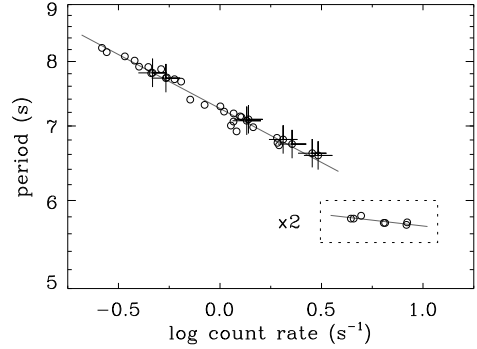}}
\caption{Period of the EUV oscillation as a function of DS count rate
during the 1996 October outburst of SS~Cyg. Points on the rising branch
of the outburst are distinguished with crosses. Grey lines are the
unweighted fits to the data: $P= 7.26\, I^{-0.097}$~s and $P=2.99\,
I^{-0.021}$~s. Points enclosed by the dotted box are plotted at twice the
observed periods.}
\end{figure}

As shown in the lower panel of Figure~1, oscillations were detected in
the {\it EUVE\/} DS count rate light curves during an interval of
approximately one week during this outburst. The oscillation was first
convincingly detected on the rising branch of the outburst at a period
of 7.81~s, it fell to 6.59~s over an interval of 4.92~hr ($Q=1.5\times
10^4$), {\it jumped\/} to 2.91~s, and then fell to 2.85~s over an
interval of 4.92~hr ($Q=3.0\times 10^5$) before observations with the DS
were terminated. When DS observations resumed 3.4 days later during the
declining branch of the outburst, the period of the oscillation was
observed to rise from 6.73~s to 8.23~s over an interval of 2.10 days
($Q=1.2\times 10^5$). When the EUV oscillation period was approximately
2.9~s and its amplitude was 25--30\%, a conservative upper limit for the
X-ray oscillation amplitude was 7\% \cite{whe03}.

\begin{figure} 
\label{fig3}
\resizebox{1.69302\columnwidth}{!}{\includegraphics{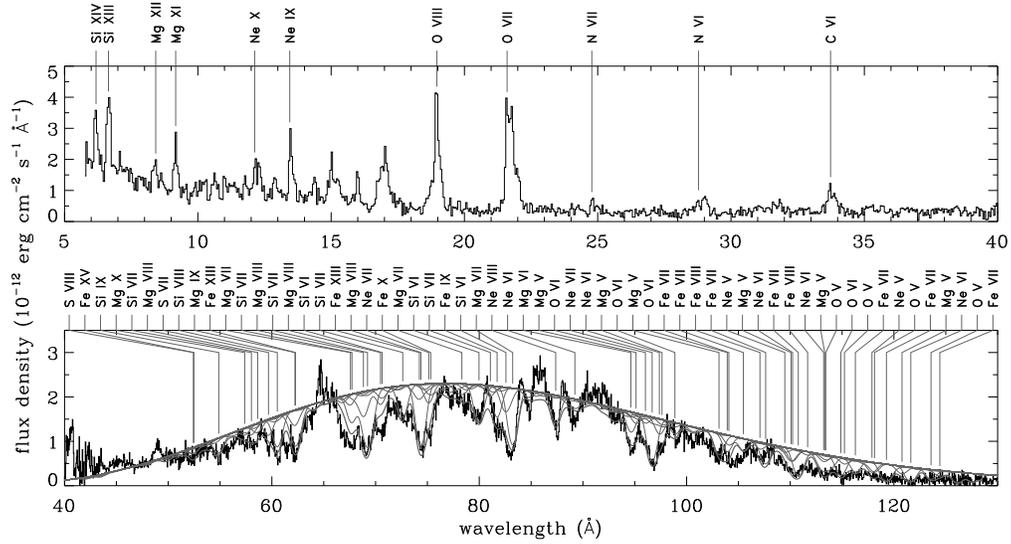}}
\caption{{\it Chandra\/} LETG spectrum of SS~Cyg obtained on 2001 January
16. Labels identify the H- and He-like lines of C, N, O, Ne, Mg, and Si
and the strongest lines in the model of the EUV spectrum. In the lower
panel, the data are shown by the  black histogram, the absorbed
blackbody continuum by the smooth gray curve, the individual ion spectra
by the gray curves, and the net model spectrum by the thick gray curve.
The strongest lines in the model are labeled.}
\end{figure}

It is clear from Figure~1 that the period of the EUV oscillation of
SS~Cyg anticorrelates with the DS count rate, being long when the count
rate is low and short when the count rate is high. To quantify this
trend, Figure~2 shows the log of the period of the oscillation as a
function of the log of the DS count rate. As in the previous figure, the
data fall into two groups: one during the early rise (distinguished with
crosses) and decline of the outburst, the other during the interval after
the frequency of the oscillation had doubled. The trends during the
early rise and decline of the outburst are the same and can be fit by a
function of the form $P=P_0\, I^{-\alpha }$ (where $I$ is the DS count
rate) with $P_0=7.26$~s and $\alpha =0.097$, consistent with the trends
observed during outbursts of SS~Cyg in 1993 August and 1994 June/July
\cite{mau96}.

Despite any direct evidence that the white dwarf in SS~Cyg is magnetic,
it is useful to consider the requirements of a magnetospheric model
to explain the period-intensity (by inference, period-$\Mdot $)
relationship observed in SS~Cyg. For a star with a dipole magnetic
field $B(r)=\mu /r^3$ (where $\mu =B(R_\star )R_\star ^3$ is the dipole
moment), the disk is truncated at a radius $r_0\propto \mu ^{4/7}
\Mdot ^{-2/7}$, hence the Keplerian frequency $\nu_{\rm K}(r_0) =
\frac{1}{2\pi } (GM_\star /r_0^3)^{1/2} \propto \Mdot ^{3/7}$. With
$\alpha = 3/7 = 0.286$, this is far ``softer'' than the relationship
observed in SS~Cyg. For a multipole magnetic field $B(r)=m_l/r^{l+2}$
(where $m_l = B(R_\star ) R_\star ^{l+2}$ is the multipole moment),
$r_0\propto m_l^{4/(4l+3)} \Mdot ^{-2/(4l+3)}$, hence $\nu_{\rm K}
(r_0)\propto \Mdot ^{3/(4l+3)}$. Under these assumptions, SS~Cyg
requires a surface magnetic field strength $B(R_\star ) \sim 0.1$--1~MG
(which is sufficiently low to be hard-to-impossible to detect directly),
and a high-order multipole field ($l = 7^{+4}_{-2}$) \cite{mau96}.

As restrictive as this is, the situation is worse during the peak of
the outburst: after the oscillation frequency doubled, the
period-intensity relationship can be fit with  $P=2.99$~s and $\alpha =
0.021$: not only did the oscillation frequency double, it's dependence
on the DS count rate became ``stiffer'' by a factor of approximately 5.
This evolution is consistent with SS~Cyg pulsating at a fundamental
period $P\gax 6.5$~s, then switching to a first harmonic and stiffening
its period-intensity (by inference, period-$\Mdot $) relationship so as
to avoid oscillating faster than $2\, P_{\rm min} \approx 5.6$~s. This
minimum period is consistent with the Keplerian period at the inner
edge of the accretion disk if the mass of the white dwarf $\Mwd\approx
1.1~\Msun $ (and the Nauenberg \cite{nau72} white dwarf mass-radius
relation applies). A secure value of the white dwarf mass is needed to
confirm this interpretation.

As shown in the upper panel of Figure~1, the McDonald Observatory $UBVR$
photometry was obtained on three consecutive nights during the early
decline of the outburst. Oscillations in the optical flux were detected
on the second and third nights with periods of 6.58~s and 6.94~s,
respectively, consistent with the corresponding EUV oscillation periods.
During these intervals, the amplitude of the EUV oscillation was 34\%,
while in $UBVR$ the amplitudes were 0.11\%, 0.07\%, 0.05\%, and 0.07\%,
respectively. On the third night, during two intervals of strictly
simultaneous optical and EUV data, the periods and phases of the
oscillations were determined to be the same within the errors. The phase
difference $\Delta \phi _0= 0.014 \pm 0.038$ implies $\Delta t= 0.10\pm
0.26$~s for $P=6.94$~s. The $3\,\sigma $ upper limit $\Delta t\le 0.88$
s corresponds to a distance $r =c\, \Delta t\le 2.6\times 10^{10}$~cm.
Assuming that the EUV oscillation originates near the white dwarf and
that the optical oscillation is formed by reprocessing of EUV flux in
the surface of the accretion disk, the delay $\Delta t=r\, (1-\sin i\,
\cos \varphi )/c$, where the binary inclination $i\approx 40^\circ $ and
$0\le\varphi\le\pi$ is the azimuthal angle from the line of sight. Then,
the distance to the reprocessing site $r=c\,\Delta t/(1-\sin i\, \cos
\varphi )\le 1.6\times 10^{10}$~cm, which is about 30 white dwarf radii
or one-third the size of the disk.

\section{Chandra LETG Observations}

Mauche \cite{mau02c, mau02} presented photometric and spectroscopic
results from a {\it Chandra\/} Low Energy Transmission Grating (LETG)
observation of SS~Cyg obtained on 2001 January 16 during the plateau
phase of a wide asymmetric outburst. The resulting spectrum, shown in
Figure~3, contains emission lines of H- and He-like C, N, O, Ne, Mg,
and Si and Fe L-shell ions in the \mbox{X-ray} band, and a
quasi-continuum extending from 40~\AA \ to 130~\AA \ in the EUV band.
The EUV spectrum can be modeled as an absorbed blackbody with flux
scattered out of the line of sight by scores of ground-state transitions
of a broad range of ions in the system's outflowing wind. The model fit
shown in the figure has the following parameters: blackbody temperature
$kT=21.5$ eV, neutral hydrogen column density $N_{\rm H}=5.0\times
10^{19}~\rm cm^{-2}$, fractional emitting area $f=4.5\times 10^{-3}$
(hence luminosity $L= 4\pi f\Rwd ^2 \sigma T^4 = 2\times 10^{33}~\rm
erg~s^{-1}$), and wind velocity $V=2500~\rm km~s^{-1}$ and mass-loss
rate $\Mdot = 3\times 10^{-11}~\rm \Msun~yr^{-1}$. The strongest lines
in the model are labeled in the figure and include
\ion{O}{5}--{\sc\uppercase\expandafter{\romannumeral 6}},
\ion{Ne}{5}--{\sc\uppercase\expandafter{\romannumeral 8}},
\ion{Mg}{5}--{\sc\uppercase\expandafter{\romannumeral 10}},
\ion{Si}{6}--{\sc\uppercase\expandafter{\romannumeral 9}},
\ion{S}{7}--{\sc\uppercase\expandafter{\romannumeral 8}},
and intermediate change states of L-shell Fe.

\begin{figure} 
\label{fig4}
\resizebox{0.88214\columnwidth}{!}{\includegraphics{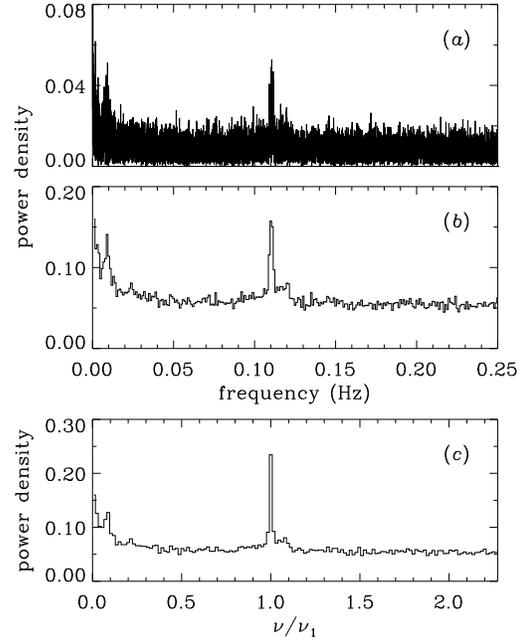}}
\caption{Power spectra of {\it Chandra\/} LETG $\lambda =42$--120 \AA \
count rate light curves of SS~Cyg in outburst. ({\it a\/}) Power spectrum
of 47 ks of data binned to 1 s time resolution. ({\it b\/}) Mean power
spectrum of 47 consecutive 1 ks light curves. ({\it c\/}) Mean power
spectrum of the 47 1 ks light curves after scaling by the varying
frequency of the $\nu_1\approx 0.11$ Hz oscillation. Note the simultaneous
presence of oscillations at $\nu_0\approx 0.0090$ Hz, $\nu_1\approx 0.11$
Hz ($\nu_0/\nu_1\approx 0.088$), and possibly a third at $\nu_2\approx
\nu_0+\nu_1\approx 0.12$ Hz.}
\end{figure}

Power spectra were calculated from background-subtracted count rate
light curves constructed from the $\pm $ first-order LETG event data.
The X-ray and EUV components of the spectrum were isolated by applying
$\lambda = 1$--42 \AA \ and $\lambda = 42$--120 \AA \ wavelength filters,
respectively. Consistent with the 1996 October {\it EUVE\/} and {\it
RXTE\/} results, the optically thick EUV component of the LETG spectrum
oscillates, while the optically thin X-ray component does not. The power
spectrum of the EUV light curve, shown in the upper panel of Figure~4,
manifests excess power at frequencies $\nu_0\approx 0.0090$ Hz and
$\nu_1\approx 0.11$ Hz, indicating the presence of oscillations at
periods $P_0\approx 110$ s and $P_1\approx 9.1$ s. To investigate these
oscillations more closely, the soft X-ray light curve was divided into
47 consecutive 1 ks intervals. Although the power spectra of these light
curves are rather noisy, the $\nu_1\approx 0.11$ Hz oscillation
typically appears as a single $\Delta\nu =0.001$ Hz peak with a
frequency in the range $\nu_1 = 0.109$--0.112 Hz. The mean of these 47
power spectra is shown in the middle panel of Figure 4. In addition to
the peaks at $\nu_0\approx 0.0090$ Hz and $\nu_1\approx 0.11$ Hz, there
appears to be a shoulder on the higher frequency peak extending to
$\nu_2\approx \nu_0+\nu_1\approx 0.12$ Hz. This feature is revealed as a
distinct peak in the power spectrum of the harder ($\lambda = 42$--70
\AA ) half of the EUV component of the LETG spectrum. In {\it EUVE\/} DS
data obtained during the 1994 June/July outburst of SS~Cyg, oscillations
were detected at $\nu_0\approx 0.012$ Hz and $\nu_1\approx 0.13$ Hz (and
its first harmonic $2\nu_1$), indicating the presence of oscillations
at periods $P_0\approx 83$ s and $P_1\approx 7.7$ s \cite{mau97c}. The
ratio $P_1/ P_0$ for the {\it Chandra\/} LETG and {\it EUVE\/} DS power
spectra is 0.088 and 0.096, respectively.

\begin{figure} 
\label{fig5}
\resizebox{0.95\columnwidth}{!}{\includegraphics{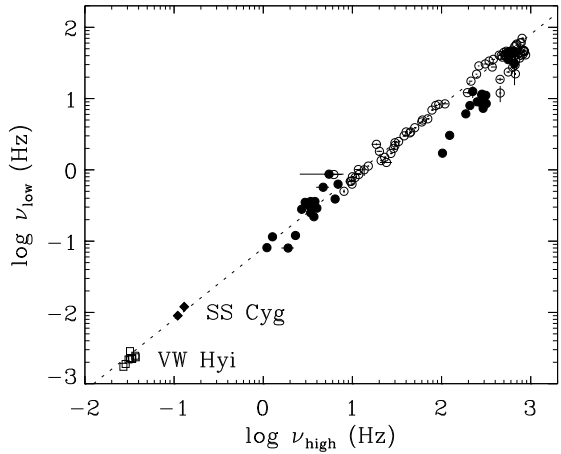}}
\caption{$\nu_{\rm high}$--$\nu_{\rm low}$ correlation for neutron
star binaries ({\it open circles}), black hole binaries ({\it filled
circles}), and the white dwarf binaries SS~Cyg ({\it filled diamonds\/})
and VW~Hyi ({\it open squares\/}). The SS~Cyg data are from Mauche
\cite{mau02}, the VW~Hyi data are from Woudt \& Warner \cite{wou02}, and
the neutron star and black hole binary data are from Belloni, Psaltis,
\& van der Klis \cite{bel02}, and were kindly supplied by T.\ Belloni.
Dotted line drawn through the points is $\nu_{\rm low} = 0.08 \,
\nu_{\rm high}$.}
\end{figure}

SS~Cyg is not the only compact binary in which multiple periodicities
have been detected. Woudt \& Warner \cite{wou02} discuss a number of
instances when multiple periodicities were detected in the optical flux
of VW~Hyi in outburst. During the decline of the 2000 February outburst,
DNOs with periods $P_{\rm DNO}=27$--37 s and QPOs with periods $P_{\rm
QPO}=400$--580 s were detected simultaneously. The ratio $P_{\rm DNO}/
P_{\rm QPO}=0.064$--0.071 is similar to the period ratios observed in
SS~Cyg. In addition, Psaltis, Belloni, \& van der Klis \cite{psa99}
showed that in five Z sources a tight correlation exists between the
``horizontal branch oscillation'' (HBO) frequency $\nu_{\rm HBO}$ and
the frequency $\nu_l$ of the lower frequency member of the pair of
``kilohertz'' (kHz) QPOs, with $\nu_{\rm HBO}/\nu_l\approx 0.12$.
Furthermore, by identifying with $\nu_{\rm HBO}$ and $\nu_l$ the
frequencies of various types of peaked noise components in atoll
sources, other neutron star binaries, and black hole binaries, they (and
subsequently Belloni, Psaltis, \& van der Klis \cite{bel02}) extended
this correlation over nearly three orders of magnitude in frequency.
Figure~5 shows this correlation and shows that the EUV data of SS~Cyg
and the optical data of VW~Hyi extend this correlation nearly two
orders of magnitude lower in frequency. This connection between the
oscillations of CVs and LMXBs has since been strengthened by Warner,
Woudt, \& Pretorius \cite{war03}, who used published, archival, and
new data to add a number of other CVs to the correlation, extending
it another order of magnitude lower in frequency (to the magnetic
disk-accreting dwarf nova GK~Per, for which $\nu_{\rm low}$ is the QPO
frequency and $\nu_{\rm high}$ is the white dwarf spin frequency).

This connection between the oscillations of CVs and LMXBs identifies the
DNOs of CVs with the kHz QPOs of LMXBs, and the QPOs of CVs with the
HBOs (or the broad noise component identified as such) of LMXBs. Note
that the frequencies of the DNOs of CVs and the kHz QPOs of neutron
star binaries are similar in that they are comparable to the Keplerian
frequency at the inner edge of the accretion disk of, respectively, a
white dwarf and neutron star: $\nu_{\rm K}\le 0.14$ Hz for a $M_\star
=1\, \Msun $ white dwarf with $r\ge R_\star=5.5\times 10^8$ cm, while
$\nu_{\rm K}\lax 1570$ Hz for a $M_\star =1.4\, \Msun $ neutron star
with $r\gax 3\, R_{\rm S}=6\, GM_\star/c^2=12.4$ km, as required by
general relativity. In addition to their frequencies, the DNOs of CVs
and the kHz QPOs of neutron star binaries are similar in that they have
relatively high coherence and high amplitudes, their frequency scales
with the inferred mass-accretion rate, and they sometimes occur in
pairs. Assuming that the same mechanisms produce the oscillations in
white dwarf, neutron star, and black hole binaries, the data exclude
the relativistic precession model and the magnetospheric and sonic-point
beat-frequency models (as well as {\it any\/} model requiring the
presence or absence of a stellar surface or magnetic field) \cite{mau02}.

\begin{figure} 
\label{fig6}
\resizebox{0.95000\columnwidth}{!}{\includegraphics{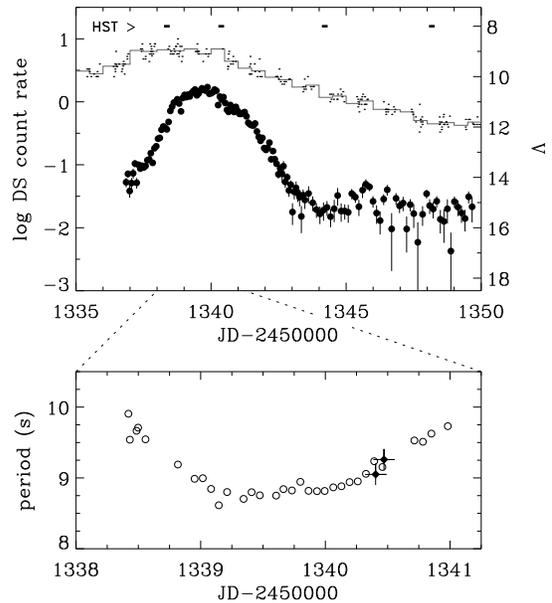}}
\caption{{\it Upper panel\/}: {\it EUVE\/} and AAVSO optical light curves
of the 1999 June outburst of SS~Cyg. DS measurements are shown by the
filled circles with error bars; individual AAVSO measurements are shown
by the small dots; half-day mean optical light curve is shown by the
histogram. Intervals of {\it HST\/} observations are indicated by the
thick bars. {\it Lower panel\/}: Oscillation period versus time. {\it
EUVE\/} DS and {\it HST\/} UV measurements are shown by the open circles
and starred filled diamonds, respectively.}
\end{figure}

\begin{figure} 
\label{fig7}
\resizebox{1.75\columnwidth}{!}{\includegraphics{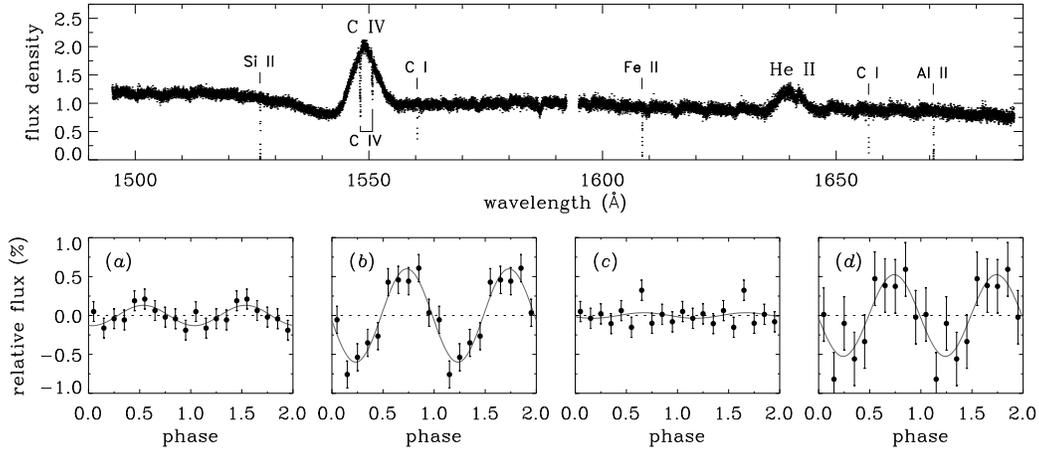}}
\caption{{\it Upper panel\/}: {\it HST\/} STIS E140H spectrum of SS~Cyg
obtained on 1999 June 10.86 UT. Labels identify the \ion{C}{4} P Cygni
profile, the \ion{He}{2} emission line, and various strong interstellar
absorption lines. Units of flux density are $10^{-11}$ erg cm$^{-2}$
s$^{-1}$ \AA $^{-1}$. {\it Lower panel\/}: Events folded on the
oscillation period $P=9.28$ s for echelle orders
({\it a\/})  1--6  ($\lambda\approx 1495$--1528 \AA ),
({\it b\/})  8--11 ($\lambda\approx 1533$--1557 \AA : \ion{C}{4}),
({\it c\/}) 13--23 ($\lambda\approx 1562$--1629 \AA ), and
({\it d\/}) 25--26 ($\lambda\approx 1634$--1648 \AA : \ion{He}{2}).}
\end{figure}

\section{HST STIS Observations}

The final campaign to be discussed is one in which {\it EUVE\/} and
{\it HST\/} Space Telescope Imaging Spectrograph (STIS) observations
were obtained during an outburst of SS~Cyg in 1999 June. The resulting
optical and EUV lights curves are shown in the upper panel of Figure~6.
Unfortunately, the outburst proved to be both short, so only two of the
{\it HST\/} visits coincided with the peak of the outburst, and weak,
so the peak DS count rate was relatively low. EUV oscillations were
detected in the DS count rate light curves during an interval of
approximately $2\frac{1}{2}$ days near the peak of the outburst. The
lower panel of Figure~6 shows the evolution of the period of the EUV
oscillation, which gradually fell from $\approx 10$~s to $\approx
8.75$~s, and then rose back to $\approx 10$~s. Four orbits of {\it
HST\/} observations were obtained during each of four visits on 1999
June 8, 10, 14, and 18. During each visit, STIS was used in TIME-TAG
mode with (1) the NUV-MAMA with the G230L grating ($\lambda\approx
1600$--2800~\AA , one orbit), (2) the FUV-MAMA with the G140L grating
($\lambda\approx 1150$--1600~\AA, one orbit), and (3) the FUV-MAMA
with the E140H echelle grating centered at $\lambda = 1598$ \AA \
($\lambda\approx 1495$--1690~\AA , two orbits). Unfortunately, it
was necessary for reasons of detector safety to use the F25NDQ
aperture/neutral density filter with the first two configurations. UV
oscillations were detected during two consecutive orbits on the early
decline from outburst in light curves produced from the echelle grating
event data ($2\times 10^7$ photons per orbit!) with periods of 9.06~s
and 9.28~s, consistent with the corresponding EUV oscillation periods.

More interesting is the spectrum of the UV oscillations. The upper
panel of Figure~7 shows the spectrum derived from the echelle grating
observation obtained on 1996 June 10.86 UT. It consists of a bright
continuum on which are superposed the \ion{C}{4} $\lambda1549$ P~Cygni
profile, the \ion{He}{2} $\lambda1640$ emission line, and various
interstellar absorption lines (the discontinuities in the spectrum every
$\approx 6$~\AA \ are due to small errors in the assumed effective areas
of the 32 echelle orders). Power spectra were calculated for light curves
of each echelle order, and excess power was detected in the neighborhood
of the \ion{C}{4} and \ion{He}{2} lines, but not in the continuum. To
determine the wavelength dependence of the oscillation while maximizing
the signal, power spectra were calculated for light curves produced from
events from echelle orders
  1--6  ($\lambda\approx 1495$--1528 \AA ),
  8--11 ($\lambda\approx 1533$--1557 \AA : \ion{C}{4}),
13--23 ($\lambda\approx 1562$--1629 \AA ), and
25--26 ($\lambda\approx 1634$--1648 \AA : \ion{He}{2}).
The lower panels of Figure~7 show the relative light curves for each of
these channels folded on the 9.28~s oscillation period. By fitting a
sinusoidal function $A +B\, \sin 2\pi (\phi -\phi _0)$ to the folded
light curves, the relative oscillation amplitudes $B/A$ were found to
be $0.13\%\pm 0.06\%$, $0.60\%\pm 0.08\%$, $0.04\%\pm 0.06\%$, and
$0.53\%\pm 0.15\%$, respectively: the oscillation was convincingly
detected in the \ion{C}{4} and \ion{He}{2} channels, only weakly (at
$2.2\sigma $) detected in the $\lambda\approx 1495$--1528 \AA \
continuum channel, and not detected in the $\lambda\approx 1562$--1629
\AA \ continuum channel. This result is in contrast to the eclipsing
dwarf nova OY~Car in superoutburst, which showed oscillations in the
UV continuum, but not in the \ion{C}{4} emission line \cite{mar98}.
The result for SS~Cyg does not exclude the possibility that the
oscillation amplitude of the UV continuum was as high as is observed
in the optical (0.05\%--0.1\%, \S 2). The possibility remains that the
optical oscillation, like the UV oscillation, is dominated by the
lines. In support of this, note that in the dwarf nova V2051 Oph in
outburst, the mean oscillation amplitude of the Balmer emission lines
is approximately twice that of the optical continuum \cite{ste01}.

How are we to understand these results? Numerous lines of evidence
identify the white dwarf/boundary layer as the primary source of the
oscillations observed in nonmagnetic, disk-accreting, high-$\Mdot $ CVs.
The pulsations of intermediate polars and the $360^\circ $ phase shifts
of eclipsing intermediate polars \cite{war72}, novalike variables
\cite{nat74, kni98}, and dwarf novae in outburst \cite{pat81} are
interpreted in terms of a model in which rotating beams of EUV/X-ray
photons are reprocessed in the surface of the accretion disk
\cite{pat79, pet80}. Warner \& Woudt \cite{war02} argue that DNOs
are caused by an equatorial belt on the surface of the white dwarf that
is first spun up and then spun down during dwarf nova outbursts, and
that QPOs are caused by vertical thickenings of the inner disk, which
alternately obscure and reflect radiation from the central source.

The \ion{He}{2} $\lambda1640$ Balmer $\alpha $ line arises from the
surface of the accretion disk and is a probe of EUV flux, as it arises
from the cascade following photoionization of \ion{He}{2} by EUV
($\lambda < 228$ \AA ) photons. The \ion{C}{4} feature, on the other
hand, is formed by scattering of $\lambda =1548$, 1550 \AA \ photons in
SS~Cyg's outflowing wind. It is something of a surprise that this line
pulsates at all, given the large volume in which it is formed. Even
more surprising is the apparent lack of a phase shift across the
\ion{C}{4} profile: the blue and red sides of the profile rise and fall
in phase. To investigate the implications of this result, we constructed
a simple model in which photons from a central source were scattered
into the line of sight by a spherically symmetric constant-velocity
wind, taking into account the light travel delays for the scattered
photons. Different pulse-phase resolved profiles were obtained for a
source with a rotating one-armed beam, a rotating two-armed beam, and
an isotropic source whose flux is sinusoidally modulated. The synthetic
profiles showed the greatest phase shift between the blue and red
sides of the profile for the rotating one-armed beam source and the
least phase shift for the modulated isotropic source, but a more
detailed investigation is required to determine if the rotating
two-armed beam source can be excluded by the data.

\section{Conclusion}

As I hope this contribution has demonstrated, the future of the study
of the rapid periodic oscillations of CVs is bright. Additional
observations of additional systems can help establish the proposed
equivalence of CV and LMXB QPOs, and will provide unique and
quantitative tests of QPO models. Data can be obtained from the ground
in the optical and from space in the UV and EUV, the system parameters
(binary inclination, white dwarf mass, radius, and rotation velocity)
can be measured, eclipse mapping in edge-on systems allows the sites of
flux modulations to be located and dissected, dwarf nova outbursts
provide a dramatic and systematic variation in the mass-accretion rate,
and diagnostic emission lines are available for study in the optical,
UV, and EUV. The growth area in this field of research in the near
future will likely be high-speed spectroscopic observations of CVs
with large optical telescopes, such as Keck and the VLT.


\begin{theacknowledgments}
I thank my collaborators J.\ Mattei, E.\ Robinson, J.\ Raymond, and
P.\ Wheatley for their contributions to this research, the AAVSO for
target-of-opportunity alerts and optical data, P.~Wheatley for
supplying the {\it RXTE\/} data shown in Fig.~1, and T.\ Belloni for
supplying the neutron star and black hole binary data shown in Fig.~5.
Support for this
work was provided in part by NASA through (1) {\it Chandra\/} Award
Number GO1-2023A issued by the Chandra X-ray Observatory Center, which
is operated by SAO for and on behalf of NASA under contract NAS8-39073
and (2) grant number GO-06545.01-95A from STScI, which is operated
by the AURA, under NASA contract NAS 5-26555. This work was performed
under the auspices of the U.S.~Department of Energy by University of
California Lawrence Livermore National Laboratory under contract
No.~W-7405-Eng-48.
\end{theacknowledgments}



\end{document}